\documentclass[]{IEEEtran}
\usepackage{cite}
\usepackage{amsmath,amssymb,amsfonts}
\usepackage{algorithm}
\usepackage{algpseudocode}
\usepackage[pdftex]{graphicx}
\usepackage{textcomp}
\usepackage{xcolor}
\usepackage{mathtools}

\ifCLASSOPTIONcompsoc
\usepackage[caption=false, font=normalsize, labelfont=sf, textfont=sf]{subfig}
\else
\usepackage[caption=false, font=footnotesize]{subfig}
\fi

\usepackage{url}	
\usepackage{siunitx}	
\usepackage{booktabs}
\usepackage{tabularx}
\newcolumntype{Y}{>{\centering\arraybackslash}X}
\usepackage{makecell}
\usepackage{multirow}
\usepackage{acro}
\usepackage{gensymb}
\acsetup{single}
\usepackage{flushend}
\usepackage{tikz}

\usepackage{soul}
\usepackage[most]{tcolorbox}

\DeclareAcronym{aoa}{
	short = AoA,
	long = angle of arrival
}

\DeclareAcronym{crlb}{
	short = CRLB,
	long = Cramer-Rao lower bound
}

\DeclareAcronym{edm}{
	short = EDM,
	long = Euclidean distance matrix
}

\DeclareAcronym{edmcp}{
	short = EDMCP,
	long = \ac{edm} completion problem
}

\DeclareAcronym{evd}{
	short = EVD,
	long = eigenvalue decomposition
}

\DeclareAcronym{evm}{
	short = EVM,
	long = error vector magnitude
}

\DeclareAcronym{ga}{
	short = GA,
	long = genetic algorithm
}

\DeclareAcronym{iot}{
	short = IoT,
	long = Internet of Things
}

\DeclareAcronym{knn}{
	short = k-NN,
	long = k-nearest-neighbor
}

\DeclareAcronym{los}{
	short = LoS,
	long = line-of-sight
}

\DeclareAcronym{lfm}{
	short = LFM,
	long = linear frequency modulation
}

\DeclareAcronym{ls}{
	short = LS,
	long = least-squares
}

\DeclareAcronym{lut}{
	short = LUT,
	long = lookup table
}

\DeclareAcronym{mds}{
	short = MDS,
	long = multidimensional scaling
}

\DeclareAcronym{nlos}{
	short = NLoS,
	long = non-\ac{los}
}

\DeclareAcronym{pps}{
	short = PPS,
	long = pulse-per-second
}

\DeclareAcronym{prf}{
	short = PRF,
	long = pulse repetition frequency
}

\DeclareAcronym{pri}{
	short = PRI,
	long = pulse repetition interval
}

\DeclareAcronym{psd}{
	short = PSD,
	long = positive semidefinite
}

\DeclareAcronym{qls}{
	short = QLS,
	long = quadratic least-squares
}

\DeclareAcronym{rf}{
	short = RF,
	long = radio frequency
}

\DeclareAcronym{rss}{
	short = RSS,
	long = received signal strength
}

\DeclareAcronym{sdr}{
	short = SDR,
	long = software-defined radio
}

\DeclareAcronym{snr}{
	short = SNR,
	long = signal-to-noise ratio
}

\DeclareAcronym{svr}{
	short = SVR,
	long = support vector regression
}

\DeclareAcronym{tdoa}{
	short = TDoA,
	long = time difference of arrival
}

\DeclareAcronym{toa}{
	short = ToA,
	long = time of arrival
}

\DeclareAcronym{tof}{
	short = ToF,
	long = time-of-flight
}

\DeclareAcronym{uhd}{
	short = UHD,
	long = \ac{usrp} Hardware Driver
}

\DeclareAcronym{usrp}{
	short = USRP,
	long = Universal Software Radio Peripheral
}

\DeclareAcronym{uwb}{
	short = UWB,
	long = ultra-wideband
}

\DeclareAcronym{wsn}{
	short = WSN,
	long = wireless sensor network
}

\DeclareMathOperator*{\argmin}{arg\,min}

\begin{document}

\title{Decentralized Localization of Distributed Antenna Array Elements Using an Evolutionary Algorithm}

\author{ Matthew J. Dula,~\IEEEmembership{Graduate~Student~Member,~IEEE}, Naim Shandi,~\IEEEmembership{Graduate~Student~Member,~IEEE},\\and Jeffrey A. Nanzer,~\IEEEmembership{Senior Member,~IEEE}
	\thanks{This work was supported in part by the Army Research Laboratory under
Cooperative Agreement Number W911NF2420011, in part by the Office of Naval Research under award \#N00014-20-1-2389, in part by the National Science Foundation under grant \#1751655, and in part by the Google Research Scholar program.}
	\thanks{	
		M. J. Dula, N. Shandi, and J. A. Nanzer are with the Department of Electrical and Computer Engineering, Michigan State University, East Lansing, MI 48824 USA (email:  dulamatt@msu.edu, shandina@msu.edu, nanzer@msu.edu).}
}
	
	\maketitle
	
\begin{abstract}

Distributed phased arrays have recently garnered interest in applications such as satellite communications and high-resolution remote sensing. High-performance coherent distributed operations such as distributed beamforming are dependent on the ability to synchronize the spatio-electrical states of the elements in the array to the order of the operational wavelength, so that coherent signal summation can be achieved at any arbitrary target destination. In this paper, we address the fundamental challenge of precise distributed array element localization to enable coherent operation, even in complex environments where the array may not be capable of directly estimating all nodal link distances. We employ a two-way time transfer technique to synchronize the nodes of the array and perform internode ranging. We implement the classical multidimensional scaling algorithm to recover a decentralized array geometry from a set of range estimates. We also establish the incomplete set of range estimates as a multivariable non-convex optimization problem, and define the differential evolution algorithm which searches the solution space to complete the set of ranges. We experimentally demonstrate wireless localization using a spectrally-sparse pulsed two-tone waveform with 40\,MHz tone separation in a laboratory environment, achieving a mean localization \ac{evm} of 0.82\,mm in an environment with an average link SNR of 34\,dB, theoretically supporting distributed beamforming operation up to 24.3\,GHz.

\end{abstract}


\acresetall

\section{Introduction}
\label{sec:intro}

\begin{figure}
	\centering
	\includegraphics[width=0.75\columnwidth]{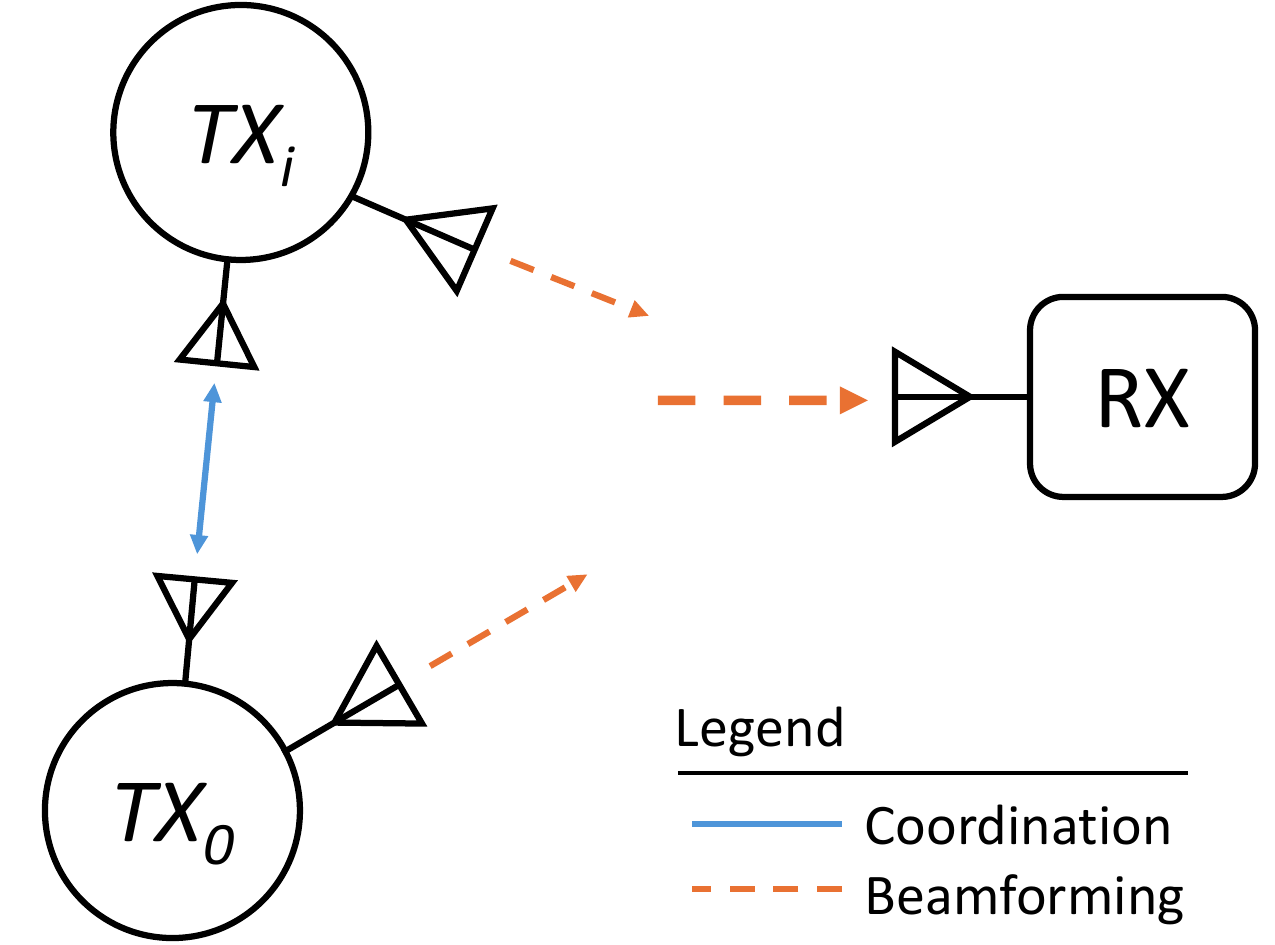}
	\caption{Distributed antenna array systems self-coordinate wirelessly in order to enable distributed beamforming, among other operations requiring phase coherence. Some of the principal challenges for reliable array operation include alignment in the system's phase, frequency, and time.}
	\label{fig:distributed-array}
\end{figure}

\IEEEPARstart{D}{istributed phased arrays} are continually advancing as a key asset for a multitude of next-generation technologies, including satellite-based 6G cellular networks~\cite{yuan2011}, imaging and detection in autonomous vehicle infrastructure~\cite{hasnain2018,tagliaferri2021}, and distributed remote sensing applications~\cite[TX08.2.3]{nasa2020tax}. Distributed antenna array systems, as conceptually shown in Fig. \ref{fig:distributed-array}, provide a number of advantages over conventional monolithic systems, including increases in adaptability and scalability, as well as resilience against single-point failures. In order to improve performance of traditional single-platform wireless systems, improvements are required to some fundamental property of the antenna; input power limits, efficiency, or system size usually must be increased. As the array size increases, the ability to improve such systems diminishes since device technologies limit advancements; however, with a distributed architecture the limitations of individual antennas or arrays can be overcome by coordinating the operation of multiple nodes, improving performance. For any specific application, it then becomes possible to achieve the necessary gain by adding nodes to the network, or by making smaller improvements on a more granular scale. Additionally, since the nodes are not necessarily fixed in space, they can be rearranged or replaced to fit task- or environment-specific constraints. The distributed paradigm also offers the benefit of fault-tolerance; if any of a set of nodes fails, the system performance would be degraded, but would remain at least partially viable, unlike in the case of monolithic systems. However, in order to achieve the level of coordination necessary for coherent operation, it is essential that the nodes of the array be tightly aligned in phase, frequency, and time, ensuring that transmitted signals from each node sum coherently at the target destination~\cite{nanzer2017tmtt}. For applications at microwave and millimeter-wave frequencies, precise localization on the order of small fractions of a wavelength is necessary to achieve high coherent gain~\cite{nanzer2021tmtt}.

Previous works have explored localization of \acp{wsn} using both range-based and range-free techniques. Range-free methods are heavily reliant on machine learning algorithms such as \ac{knn} estimation and neural networks \cite{zafari2015journal}, and while they can be very accurate, they are very computationally impractical for outdoor environments due to the arbitrarily large training datasets needed for precise estimation \cite{jovanovic2023}. Some of the most prominent range-based solutions include variants of \ac{rss}, \ac{toa}, \ac{tdoa}, and \ac{aoa} estimation algorithms, but these usually require prior knowledge of anchor locations or an \ac{uwb} implementation to achieve sub-centimeter localization accuracy \cite{jovanovic2018,paul2009,barsocchi2009,bisio2014,loyez2015,kravets2024}. Moreover, range-based approaches to the \ac{wsn} positioning problem can be classified in two categories based on their hierarchical structure: centralized solutions, where one node or cluster of nodes acts as a primary or anchor point from which all other positions are established \cite{alsamhi2019,bhat2022}; and decentralized solutions, which do not rely on any single node for positional information. Centralized approaches are often preferable for applications with fixed anchor nodes \cite{deutschmann2024,li2020basestation}, or where low range resolutions are acceptable, such as detection and surveillance \cite{zhang2022} or \ac{iot} \cite{zafari2019survey}. Despite the amount of research dedicated to localization, much of the focus for high-accuracy applications has been on indoor systems, where the environment can be heavily constrained. In most cases, simple geometric equations can be implemented to position nodes based on fixed anchors. However, centralized schemes are prone to large propagation of error, as nodes furthest from the primary node either rely on low \ac{snr} range estimation from the primary node, or estimate their range from secondary nodes, whose positions themselves are only an estimate. This lack of reliance on a primary node also provides an increase in fault tolerance for decentralized approaches in comparison to centralized alternatives. Therefore, in applications such as adaptable and scalable cooperative \ac{rf} beamforming where resilience and element localization accuracy are essential for reliable performance, it is desirable to develop range-based decentralized localization methods.

In this paper, we present a new method for high-precision decentralized wireless localization of distributed phased array elements that can be used for partially-connected network topologies, such as those in non-line-of-sight (NLoS) scenarios. By using a spectrally-sparse two-tone waveform in a two-way time transfer method, we are able to minimize the variance on our internode range estimation, regardless of noise in the environment or operational bandwidth \cite{jason2023tmtt}. We use a \ac{sdr} architecture, and combine the ranging technique with a genetic algorithm based on multidimensional scaling applied to the resulting \ac{edm} to experimentally demonstrate a mean localization accuracy of \SI{0.82}{\milli\meter} from a minimally-connected six node topology using a two-tone waveform with a tone separation of \SI{40}{\mega\hertz}. In previous works, we have demonstrated sub-millimeter internode range estimation accuracy using the two-way time transfer method \cite{anton2020mwcl} and sub-centimeter precision localization by combining the time transfer method with basic geometric calculations \cite{ahona2023eurad}. However, the localization technique in \cite{ahona2023eurad} relied upon a central primary node, as well as complete knowledge of internode range estimate information. We have previously shown the capability for decentralized localization from range estimates \cite{dula2024wamicon}, but this method also required the entire set of range estimates. Due to network or hardware constraints of large distributed systems, the connectivity in an array may be inherently limited as it may not be possible for every node to have a \ac{los} path or sufficient \ac{snr} with reference to every other node, in order to obtain the full set of range information. Therefore, in applications requiring a large array, it may be necessary to localize the system in such a way that minimizes the amount of required information. Here, we employ the two-way time transfer ranging technique using the two-tone waveform to maximize delay estimation accuracy. We introduce the \ac{mds} function for recovering array geometry from distance estimates. We also explain the concept of connectivitiy for distributed systems and develop an expression for minimal connectivity required for localization before describing our \ac{ga} which is used to complete the set of ranges in the presence of missing internode range information. Then, we present a performance evaluation for the algorithm based on measured data, analyzing the effect of connectivity on localization and demonstrating its robustness.

The rest of this paper is organized as follows. In Section \ref{sec:time}, we describe the time model and two-way time transfer process used for synchronization, the design of the waveform for optimal delay/range estimation accuracy, and the range estimation and refinement process. In Section \ref{sec:localization}, we present our network localization algorithm, including multidimensional scaling and the implemented genetic algorithm, along with its associated cost function. In Section \ref{sec:sim}, we show simulation results of the algorithm, evaluating the performance against several parameters. Finally, in Section \ref{sec:exp}, we discuss the hardware setup of the system, as well as the results of the algorithm applied to experimental measurements over a range of varying network topologies and ranging precisions.

\section{High-Accuracy Internode Ranging for Distributed Elements}
\label{sec:time}

The ability to transmit signals from a distributed architecture such that they sum coherently at an intended destination is dependent on appropriate coordination of the electrical states of the nodes. This coordination is tightly bound to the relative position of nodes in the array, and thus it is important to be able to accurately estimate the distance between nodes. In order to have a 90\% probability of 90\% of the transmitted power being available at the target, it has been shown that the standard deviation of the ranging error, $\sigma_d$, must be below $\lambda_c/15$, where $\lambda_c$ is the wavelength of the beamforming carrier frequency \cite{nanzer2021tmtt}. Thus, in order to enable beamforming capabilities at microwave and mm-wave frequencies, it is necessary to produce location estimates with sub-millimeter accuracy.

\subsection{Two-Way Time Transfer}
\label{sub_sec:t_transfer}

\begin{figure}
	\centering
	\includegraphics[width=1\columnwidth]{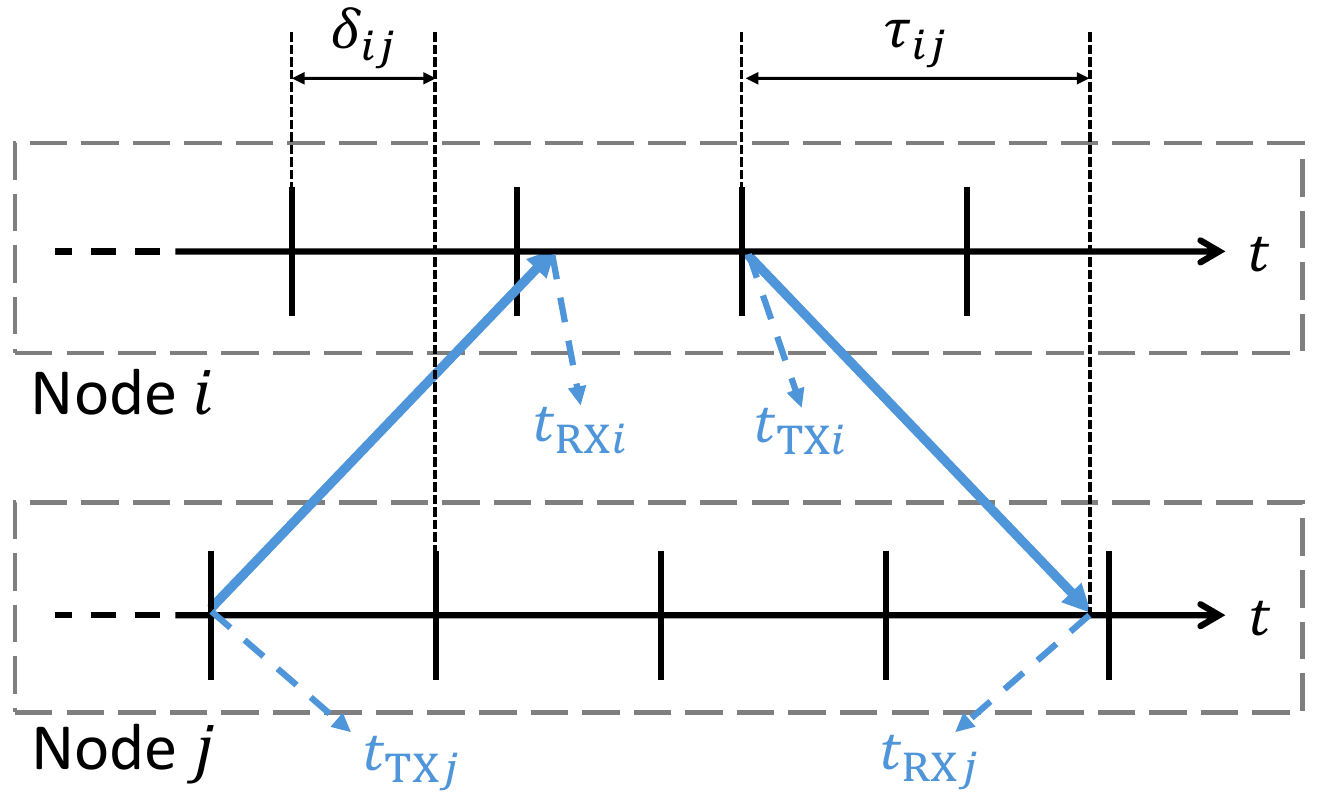}
	\caption{The two-way time-transfer design. Node $j$ transmits a timing waveform, and after reception, node $i$ retransmits this waveform back; all four timestamps are recorded and used to calculate clock offset and internode \ac{tof}.}
	\label{fig:time-transfer}
\end{figure}

Estimation of the distance between nodes is proportional to estimating the delay of a signal exchanged between the two nodes by a factor of the speed of light, thus we base our approach on high-accuracy delay estimation using a two-way time transfer method. For a distributed system consisting of $N$ nodes, the local clock time for all nodes can be modeled as a time-varying offset from a global true time, $t$, by the following:
\begin{equation}
	\label{eq:t_model}
	t_n(t) = t + \epsilon_n(t)\ \ \text{for} \ n = 0,1,...,N-1
\end{equation}
where $\epsilon_n(t)$ represents the time-varying bias of node $n$ from true time, and can be represented as the sum of a clock offset term, referred to in Fig. \ref{fig:time-transfer} as $\delta_{ij}$, and a noise term. While the clock offset term is time-variant in general due to oscillator drift and frequency offsets unique to each node, we assume for the purpose of this paper that the bias term is constant because the link is quasi-static and reciprocal throughout the synchronization epoch. In a two-way time transfer architecture as shown in Fig. \ref{fig:time-transfer}, each node pair sends a time estimation waveform in each direction. The apparent \ac{tof}, $\hat{\tau}_{ij}$, of the timing pulse from node $i$ to node $j$ is given by
\begin{equation}
	\label{eq:apparent_tof}
	\hat{\tau}_{ij} = t_{\text{RX}j} - t_{\text{TX}i},
\end{equation}
where $t_{\text{RX}j}$ and $t_{\text{TX}i}$ represent the local times of reception at node $j$ and transmission at node $i$, respectively. However, this apparent \ac{tof} is biased by the relative offset between the two local clocks; if we assume that the link is symmetric during the synchronization epoch, this can be resolved simply by taking the average of the apparent times-of-travel in each direction, thus eliminating the bias and yielding the true \ac{tof},
\begin{equation}
	\label{eq:true_tof}
	\tau_{ij} = \frac{(t_{\text{RX}j} - t_{\text{TX}i}) + (t_{\text{RX}i} - t_{\text{TX}j})}{2}.
\end{equation}

Another important note is that $\tau_{ij}$ is a summation of static \ac{rf} delays in the system hardware, \ac{tof} between antennas, and additional delay factors due to multipath effects in the environment. In this paper, we consider the scenario in which the elements and environment are immobile, so we treat the hardware delays and multipath factor as static delays that can be deduced under the assumption that the delay estimator is unbiased. If the direct path distance between antennas is known in a calibration setup, the delay factors caused by internal hardware and multipath effects can be calculated by subtracting the true free space \ac{tof} from the mean of the delay estimates (assuming a large enough sample size). In the event that an array is of size $N>2$ nodes, this will result in an $N\times N$ matrix of delay calibration values for the static system. After these static delays are determined the nodes may be moved. 

The two-way time transfer technique has also been utilized to synchronize the system to the level of picoseconds \cite{jason2023tmtt,naim2024tcom}, but because the technique simultaneously estimates clock offset $\delta_{ij}$ and \ac{tof} $\tau_{ij}$, high-precision synchronization by correcting the clock offsets is not necessary for range estimation and will not be discussed in this paper.

\subsection{Theoretical Ranging Precision Bound}
\label{sub_sec:wfm_design}

For such a technique requiring time delay estimation, the precision of the estimate is dependent on several variables, which can be described by the \ac{crlb} for range estimation. The \ac{crlb} is described in \cite[Chapter 7.1.3]{richards2014rsp} and derived for range estimation in \cite{nanzer2016tap} as
\begin{equation}
	\label{eq:crlb}
	\mathrm{var}(\hat{d}-d) = \sigma_\mathrm{d}^2\ge\frac{c^2}{2\zeta_f^2 (E_\mathrm{s}/N_0)},
\end{equation}
where $\sigma_d^2$ is the range estimation error variance, $c$ is the speed of light, $\zeta_f^2$ is the mean-squared bandwidth of the time transfer waveform, and $E_s/N_0$ is the signal energy to noise power spectral density ratio, which can be expressed as the product of the pulse duration $\tau_p$, pre-processed \ac{snr}, and noise bandwidth (equivalent to the sample rate $f_s$ for a complex sampler, which is what we will consider for this paper because of our selection of \acp{sdr} in the experimental setup):
\begin{equation}
	\label{eq:esno}
	\frac{E_\mathrm{s}}{N_0} = \tau_\mathrm{p}\cdot\mathrm{SNR}\cdot f_\mathrm{s}.
\end{equation}
Therefore, the variance of range estimation error can be reduced simply by arbitrarily increasing the integration window or transmit power. However, a more interesting method with which to accomplish higher precision is the design of a waveform which maximizes the mean-squared bandwidth within a given maximum spectral range. In \cite{nanzer2016tap}, it has been shown that $\zeta_f^2$ can be maximized by concentrating the spectral energy of a signal into the sides of the occupied bandwidth, thereby creating a waveform comprised of two discrete tones. For the two-tone waveform, the mean-squared bandwidth is given by
\begin{equation}
	\label{eq:tt_msb}
	\zeta_f^2 = (\pi \cdot \mathrm{B})^2,
\end{equation}
where $\mathrm{B}$ is the bandwidth, or tone separation of the disparate frequencies. Combining \eqref{eq:crlb}, \eqref{eq:esno}, and \eqref{eq:tt_msb} yields the \ac{crlb} for range estimation variance using a two-tone waveform:
\begin{equation}
	\label{eq:tt_crlb}
	\sigma_\mathrm{d}^2 \ge \frac{c^2}{2(\pi \cdot \mathrm{B})^2\cdot \tau_\mathrm{p}\cdot\mathrm{SNR}\cdot f_s}.
\end{equation}
Previous works have demonstrated the improvement of precision provided by this waveform over more traditional, less spectrally-sparse waveform designs, both theoretically and experimentally \cite{jason2023tmtt,anton2020mwcl}.

\subsection{Range Estimation and Refinement}
\label{sub_sec:ranging}

Using the two-way time transfer technique described previously and shown in Fig. \ref{fig:time-transfer}, the transmit times from each node are known very accurately because they invariably occur at a clock edge where the timing error is dependent on device hardware, but we will consider it negligible in comparison to receive time estimation. The receive times are of course not held to this constraint, and as such can occur at any point between clock edges. For the experimental setup to be described in Section \ref{sec:exp} with a sample rate of \SI{200}{\mega Sa/\second}, this can result in a significant timing error of up to \SI{5}{\nano\second}, corresponding to a range estimation error of approximately \SI{1.5}{\meter}. To prevent this, we have implemented a matched filter solution to the refinement process; the output of the continuous matched filter, $s_\text{\tiny{MF}}(t)$, is 
\begin{equation}
	\label{eq:cont_mf}
	s_\text{\tiny{MF}}(t) = s_\text{\tiny{RX}}(t)\circledast s_\text{\tiny{TX}}^*(-t)
\end{equation}
where $s_\text{\tiny{RX}}$ represents the received waveform, $s_\text{\tiny{TX}}$ is the ideal or transmitted time-transfer waveform, $\circledast$ is the convolution operator, and $(.)^*$ represents complex conjugation \cite{richards2014rsp}. For this continuous matched filter, the output will be maximized at the true time delay of the beginning of the received waveform. However, for the discrete matched filter, where the output is modeled as
\begin{equation}
	\label{eq:disc_mf}
	s_\text{\tiny{MF}}[n] = s_\text{\tiny{RX}}[n]\circledast s_\text{\tiny{TX}}^*[-n],
\end{equation}
the true time delay will almost surely fall between two samples, so we have implemented \ac{qls} interpolation \cite{richards2014rsp}. The \ac{qls} technique introduces a deterministic bias that is directly related to the \ac{qls} oversampling ratio, or the factor by which the sampling frequency exceeds the Nyquist rate. This bias can be corrected by implementing a \ac{lut}, where the deterministic bias factors are computed prior to delay estimation and stored for fast correction of errors based on the fractional delay bin \cite{moddemeijer1991}. Other interpolation methods with higher accuracy have been implemented based on \ac{ls} techniques related to sinc functions and direct matched filter outputs \cite{serge2022mwcl}, but we have chosen \ac{qls} with an \ac{lut} for its simplicity, speed, and low computational complexity.

\section{Decentralized Localization Algorithm}
\label{sec:localization}

\subsection{Classical Multidimensional Scaling}
\label{sub_sec:mds}

Given a set of points residing in an $m$-dimensional space, which are the columns of $\textbf{X}~=~[\boldsymbol{x}_0,\boldsymbol{x}_1,...,\boldsymbol{x}_{N-1}],\ \boldsymbol{x}_i \in \mathbb{R}^m$, it is simple to calculate the general Euclidean distance between two points as the magnitude of the difference between their position vectors. Equivalently, the squared-distance can be expressed as
\begin{equation}
	\label{eq:dist_calc}
	d_{ij} = ||\boldsymbol{x}_i - \boldsymbol{x}_j||^2
\end{equation}
where $||.||$ represents the Frobenius norm. By expanding this norm and ascribing the entries to $\textbf{D}$ such that $\textbf{D} = [d_{ij}]$, we are able to define the \ac{edm} for $\textbf{X}$,
\begin{equation}
	\label{eq:edm_assembly}
	\begin{aligned}
	\mathrm{edm}(\textbf{X}) &\vcentcolon= \textbf{1}\,\mathrm{diag}(\textbf{X}^\top\textbf{X})^\top - 2\textbf{X}^\top\textbf{X} + \mathrm{diag}(\textbf{X}^\top\textbf{X})\,\textbf{1}^\top\\
	&\vcentcolon= \textbf{D}
	\end{aligned}
\end{equation}
where $\textbf{1}$ represents the vector of ones, and $\mathrm{diag}(.)$ represents the vector containing the diagonal entries. From \eqref{eq:edm_assembly}, it is evident that the \ac{edm} for any set of points $\textbf{X}$ is not only a function of $\textbf{X}$, but can also be expressed as a function of the product matrix, $\textbf{X}^\top\textbf{X}$. From this forward problem, it is possible to then solve the inverse problem of recovering a set of points in a coordinate system from an \ac{edm}; while it is possible to preserve pairwise distances, the absolute position is inherently lost in the encoding process of converting position vectors to distance scalars. It has been shown that the relative positional geometry for a set of points can be reconstructed from the \ac{edm} through the process of \ac{mds} \cite{torgerson1952}. First, a product matrix $\textbf{X}^\top\textbf{X}$ corresponding to the relative geometry of $\textbf{X}$ from the \ac{edm} of the point set can be constructed as
\begin{equation}
	\label{eq:gram_assembly}
	\textbf{X}^\top\textbf{X} = -\frac{1}{2}\left( \textbf{I} - \textbf{1}\textbf{s}^\top\right) \textbf{D}\left( \textbf{I} - \textbf{s}\textbf{1}^\top\right),
\end{equation}
where $\textbf{I}$ represents the $n\times n$ identity matrix. Additionally, $\textbf{X}^\top\textbf{X}$ will be \ac{psd} for any $\textbf{s}$ such that $\textbf{s}^\top\textbf{1} = 1$ and $\textbf{s}^\top\textbf{D} \neq 0$ according to \cite{gower1982}, and will be of rank $m$. By choosing $\textbf{s} = \frac{1}{N}\textbf{1}$, the point set $\textbf{X}$ will place the centroid of the points at the origin of the local coordinate system; other centering factors can be chosen depending on the system application, but for the sake of error evaluation and simple ambiguity resolution, we have chosen to rely on geometric centering. By performing \ac{evd} on the product, we obtain
\begin{equation}
	\label{eq:evd}
	\textbf{X}^\top\textbf{X} = \textbf{U}\boldsymbol{\Lambda}\textbf{U}^\top
\end{equation}
\begin{algorithm}[t]
	\caption{Classical Multidimensional Scaling \cite{dokmanic2015}}
	\label{alg:mds}
	\begin{algorithmic}[1]
	\State \textbf{function} ClassicalMDS($\boldsymbol{D},m$)
	\State $\quad \ \textbf{S} \leftarrow \textbf{I} - \frac{1}{N}\textbf{1}\textbf{1}^\top$
	\State $\quad \ \textbf{P} \leftarrow -\frac{1}{2}\textbf{S}\textbf{D}\textbf{S}$
	\State $\quad \ [\lambda_1,...,\lambda_N],\textbf{U} \leftarrow \text{EVD}\left(\textbf{P}\right)$
	\State $\quad \ \textbf{return} \ \left[ \mathrm{diag}\left( \sqrt{\lambda_1},...,\sqrt{\lambda_m}\right) , \ \boldsymbol{0}_{m\times (N-m)}\right]\textbf{U}^\top$
	\end{algorithmic}
\end{algorithm}
where $\textbf{U}$ is the matrix constructed of orthonormal eigenvectors and $\boldsymbol{\Lambda}$ is the diagonal matrix of corresponding eigenvalues; we will assume it is sorted as $|\lambda_1| \geq |\lambda_2| \geq ... \geq |\lambda_N|$ for the remainder of this paper. Since the product is \ac{psd} and of rank $m$, all $\lambda_i$ will be non-negative (zero for $i>m\,$), so we can reconstruct our geometrically-centered point set as
\begin{equation}
	\label{eq:pts_true}
	\textbf{X} = \left[ \mathrm{diag}\left( \sqrt{\lambda_1},...,\sqrt{\lambda_m}\right) , \ \boldsymbol{0}_{m\times (N-m)}\right]\textbf{U}^\top.
\end{equation}

The steps of multidimensional scaling are outlined explicitly in Algorithm 1. One of the more important properties of the \ac{mds} technique is that it inherently minimizes the norm of the difference between the input points and the target points, so that it still performs well under slight perturbations or noise artifacts \cite{dokmanic2015,keshavan2009noisy}. This is essential considering that practical systems provide ranging estimates rather than the true internode ranges.

\subsection{Connectivity and Completability}
\label{sub_sec:connectivity}

\begin{figure}
	\centering
	\includegraphics[width=1\columnwidth]{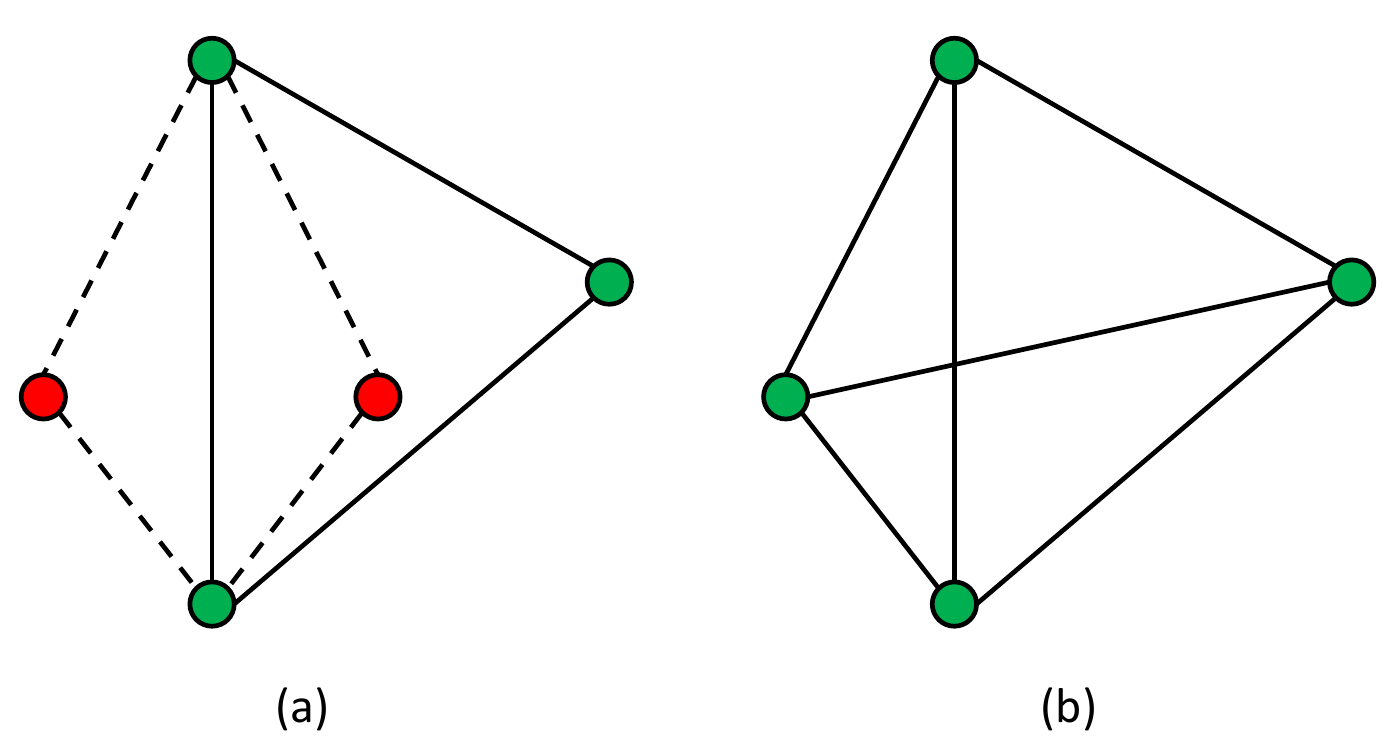}
	\caption{Illustration of the robust quadrilateral concept. In (a) the quadrilateral can satisfy distance measurements in two geometries due to the missing connection. The quadrilateral is completed in (b), resolving ambiguity.}
	\label{fig:robust-quad}
\end{figure}

In many practical settings, exact knowledge (or even estimates) of internode distances may not be available for all node pairs. This will result in missing entries to the observed \ac{edm} $\tilde{\textbf{D}}$, which must be resolved in order to localize the points with high precision; this problem is known as the \ac{edmcp}. One of the fundamental concepts to understand before attempting to solve the \ac{edmcp} is the connectivity $c$ of the array, defined as the ratio of the number of available distance estimates, referred to here as edges, to the total number of possible edges, which varies between 0 and 1. For an undirected network of $N$ nodes, the maximum number of edges is
\begin{equation}
	\label{eq:max_e}
	E_\mathrm{max} = \frac{1}{2}N(N-1).
\end{equation}
Localization for large networks, specifically $N>4$, can be based on the concept of robust completed quadrilaterals because of their inherent resilience against position ambiguities caused by positional flips, such as those shown in Fig. \ref{fig:robust-quad} \cite{rus2004}. Once a robust quadrilateral is established, further nodes can be added to the system as long as they are connected to at least $m+1$ previously-resolved nodes of the array. This constraint can be simply understood by method of intersection in the 2-dimensional case: when one distance is known, the possible positions of the new node are defined by the circle on $\mathbb{R}^2$ with a radius corresponding to the estimate. When two distances are known, the possible positions are the intersecting points of the two circles, and when three distances are known, the final distance estimate will resolve the two-point ambiguity.

With this $m+1$ constraint, we can define the minimum number of known edges for matrix completability to be a sum of the six edges in the robust quadrilateral, and $m+1$ edges for nodes beyond the fourth. Thus, the expression for the 2-dimensional case considered in this paper will be
\begin{equation}
	\label{eq:min_e}
	\begin{aligned}
	E_\mathrm{min} &= 6 + 3(N-4)\\
	&= 3N - 6.
	\end{aligned}
\end{equation}
Combining \eqref{eq:max_e} and \eqref{eq:min_e}, we obtain the following inequality for the possible values of $c$ in an array of $N$ nodes:
\begin{equation}
	\label{eq:c_defined}
	0 < c_\mathrm{min} = \frac{E_\mathrm{min}}{E_\mathrm{max}} = \frac{6(N-2)}{N(N-1)} \leq c \leq 1.
\end{equation}
It is worth noting that in practical scenarios, it is feasible to have connectivity values below $c_\mathrm{min}$, or an adjacency matrix structured such that $c$ exceeds $c_\mathrm{min}$ but contains edges such that some node(s) may not be resolvable. For example, an array of six nodes can lose up to three edges, but if all three lost edges would be connected to the same node, then the node will not be resolvable as it would then only have two remaining edges. In this event, it may be worth removing such a node, to avoid incoherent destructive interference at a beamforming destination for example. For all experiments and simulations in this paper, adjacency matrices are developed in such a way as to guarantee structural completability.

\subsection{Differential Evolution for Localization}
\label{sub_sec:de_alg}

One potential pitfall of the classical \ac{mds} approach, particularly for large array sizes, is that in treating the array network as a graph, all edge information must be known in order to localize the elements of the array with high precision. In a practical scenario, this may not be feasible for several reasons, including but not limited to \ac{los} obstructions, hardware limitations, or poorly-performing links (due to low \ac{snr}, for example). Because of this, it becomes necessary to solve the resulting \ac{edmcp}. We aim to find the coordinates whose resultant \ac{edm} best matches the observations. This can be done by minimizing the cost function~\cite{dokmanic2015}
\begin{align}
	\label{eq:cost}
	F(\textbf{X}) \vcentcolon &= \frac{1}{2}\left\lVert \textbf{W}\circ\left[ \tilde{\textbf{D}} - \text{edm}\left(\textbf{X}\right)\right]\right\rVert^2 \\
	\label{eq:cost_soln}
	\hat{\textbf{X}} &= \argmin_{\textbf{X}} F(\textbf{X})
\end{align}
where $\textbf{W}$ is the $N\times N$ adjacency matrix with ones and zeros representing the connected or disconnected links, respectively; $\tilde{\textbf{D}}$ is the partially-completed \ac{edm} based on observations; $\hat{\textbf{X}}$ is the best-fit coordinate matrix; and $\circ$ is the element-wise matrix multiplication operator. In general, the cost function described in \eqref{eq:cost} is non-convex, and depending on the size and sparseness of the array, the true global minimum can be very difficult to find. While several techniques for solving the \ac{edmcp} involving gradient descent and \ac{ls} methods have been explored \cite{fang2012,alfakih1999}, these are inherently susceptible to encountering local minima and failing to converge to the true optimum, as they rely on gradient or local information, both of which are strongly influenced by the initialization of the algorithm \cite{cetin1993}.

To overcome this challenge, we have implemented a \ac{ga}-based approach using a Python-based multi-objective evolutionary optimization library \cite{pymoo}. One of the most significant benefits of this approach is that within a bounded space, the algorithm is able to explore a broad search area by evaluating a diverse population of prospective solutions for fitness \cite{haupt2004}. Within each population is a number of individuals representing proposed solutions; in our case, each individual is a vector whose entries correspond to the ``missing'' entries of the \ac{edm} estimate. This kind of evolutionary modeling is based on the concept of natural selection; after each individual's cost has been evaluated as described in Algorithm \ref{alg:ga_loc}, a certain percentage of the best-performing individuals are selected as ``parents'' for numerical crossover and mutation, while the rest are discarded and new individuals are randomly initialized in their place. From each generation, the individual with the lowest cost is referred to as the ``result''. These steps are then repeated for any desired number of generations, or until a convergence criteria is met. For all simulation and measurement to be discussed, the criteria for convergence used is a maximum change in the objective space of $10^{-6}$.

\section{Algorithm Simulation Performance}
\label{sec:sim}

\subsection{Array \ac{snr} Characterization}
\label{sub_sec:snr_char}

In order to appropriately reference \ac{snr}, it is desirable to have a characteristic value that models some average statistic of all link \acp{snr} in the system. Because of the inverse-square relationship between distance and linear \ac{snr} for a one-way traveling signal, it can be deduced that the \ac{snr} for between two nodes $i$ and $j$ is inversely proportional to their squared distance:
\begin{equation}
	\label{eq:link_snr_def}
	\mathrm{SNR}_{ij} \propto \frac{1}{d_{ij}^2}.
\end{equation}
Taking the reciprocal of each side, we obtain
\begin{equation}
	\label{eq:recip_link_snr}
	\frac{1}{\mathrm{SNR}_{ij}} \propto d_{ij}^2,
\end{equation}
\begin{algorithm}[t]
	\caption{Genetic Localization}
	\label{alg:ga_loc}
	\begin{algorithmic}[1]
	\State \textbf{function} Localize($\tilde{\textbf{D}},\textbf{W},m$)
	\State $\quad \ $\textit{/* Initialize population */}
	\State $\quad \ $\textbf{repeat}:
	\State $\quad \ \quad \ $\textbf{for} $\boldsymbol{p}$ \textbf{in} population:
	\State $\quad \ \quad \ \quad \ $cost[$\boldsymbol{p}$] = EvaluateCost($\boldsymbol{p}$)
	\State $\quad \ \quad \ $\textbf{end}
	\State $\quad \ \quad \ \boldsymbol{p}_\mathrm{result} \leftarrow \mathrm{arg}\,\mathrm{min}(\text{cost})$
	\State $\quad \ \quad \ $\textit{/* Crossover and mutation for next generation */}
	\State $\quad \ $\textbf{until} convergence \textbf{or} MaxGenerations
	\State $\quad \ \textbf{D}_\mathrm{result} \leftarrow \tilde{\textbf{D}}$ completed with $\boldsymbol{p}_\mathrm{result}$ values
	\State $\quad \ \textbf{return} \ \mathrm{ClassicalMDS}(\textbf{D}_{\mathrm{result}},m)$
	\item[]
	\State \textbf{function} EvaluateCost($\boldsymbol{p}$)
	\State $\quad \ \textbf{D}_\text{init} \leftarrow \tilde{\textbf{D}}$ completed with $\boldsymbol{p}$ values
	\State $\quad \ \textbf{X}_\text{init} \leftarrow \text{ClassicalMDS}(\textbf{D}_{\text{init}},m)$
	\State $\quad \ \textbf{return} \ \frac{1}{2}\left\lVert \textbf{W}\circ [\tilde{\textbf{D}} - \text{edm}(\textbf{X}_\text{init})]\right\rVert^2$
	\end{algorithmic}
\end{algorithm}
and we can then take the arithmetic mean to acquire the mean-squared distance on the right and the mean of the reciprocal \acp{snr} on the left:
\begin{equation}
	\label{eq:recip_mean_snr}
	\frac{2}{N(N-1)} \sum_{i < j} \frac{1}{\mathrm{SNR}_{ij}} \propto \bar{d^2}
\end{equation}
where $\bar{d^2}$ is the mean-squared internode distance. Again taking the reciprocal of each side, we obtain the definition of the harmonic mean (reciprocal of the arithmetic mean of the reciprocals) of the linear \acp{snr}, denoted by $\bar{\mathrm{SNR}}_\mathrm{h}$ on the left-hand side:
\begin{equation}
	\label{eq:mean_relation}
	\bar{\mathrm{SNR}}_\mathrm{h} \propto \frac{1}{\bar{d^2}}.
\end{equation}
By combining the proportional relations in \eqref{eq:link_snr_def} and \eqref{eq:mean_relation}, we obtain the following relation between link \ac{snr}, characteristic (or harmonic mean) \ac{snr}, and distance:
\begin{equation}
	\label{eq:link_snr_final}
	\mathrm{SNR}_{ij} = \bar{\mathrm{SNR}}_\mathrm{h}\cdot\left(\frac{\bar{d^2}}{d_{ij}^2}\right).
\end{equation}
Note that the harmonic mean relationship defined above is also used to characterize the experimental average, discussed in Section \ref{sec:exp}.

\subsection{Monte Carlo Simulations}
\label{sub_sec:mc_sims}

\begin{figure}
	\centering
	\includegraphics[width=1\columnwidth]{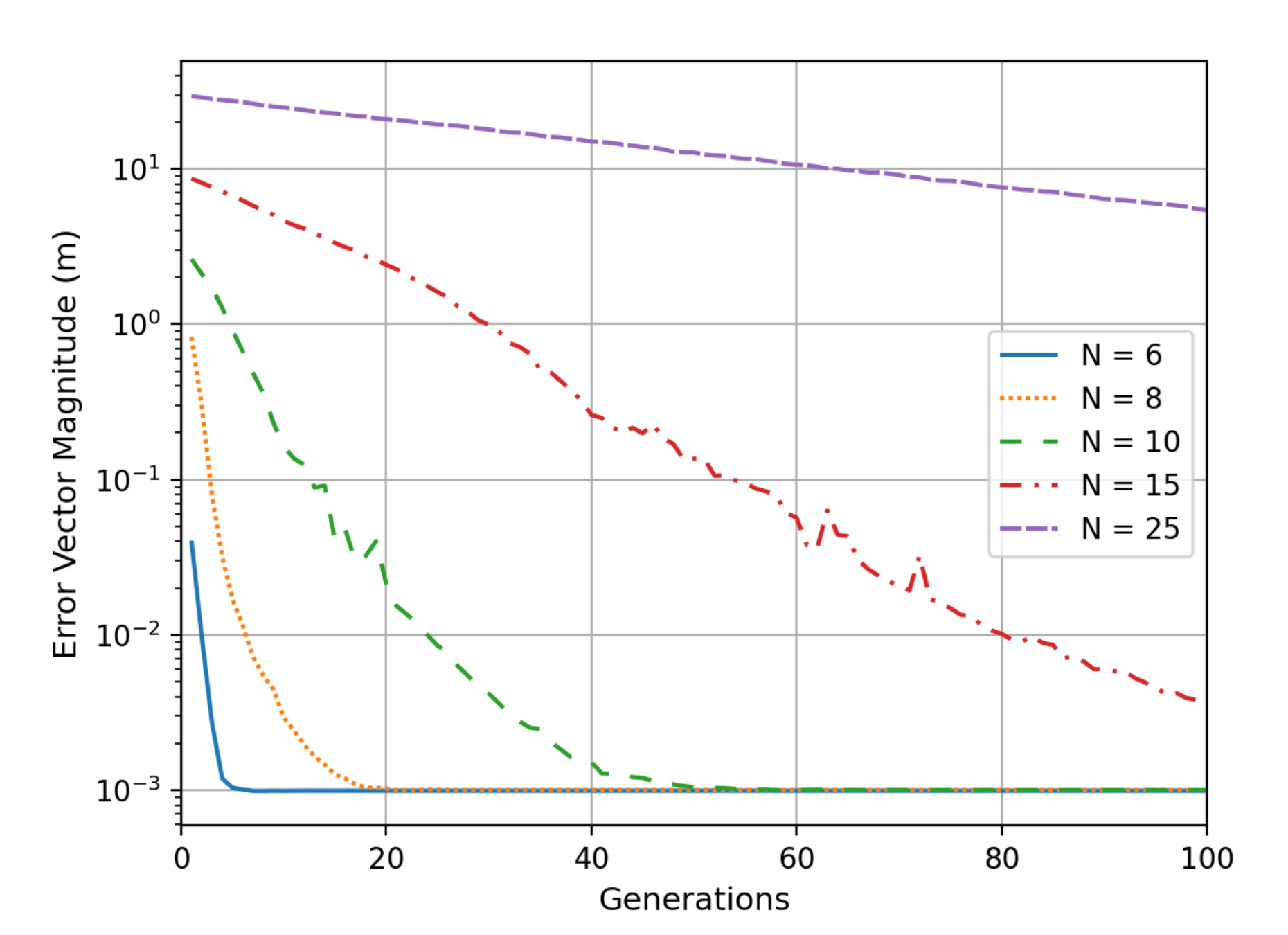}
	\caption{Simulated average algorithm localization accuracy (\ac{evm}) convergence plotted for multiple values of network connectivity. \ac{edm} estimates are generated with an array connectivity of $c=0.9$ and a time-transfer bandwidth of \SI{40}{\mega\hertz}.}
	\label{fig:sim_evm_vs_n}
\end{figure}
\begin{figure}
	\centering
	\includegraphics[width=1\columnwidth]{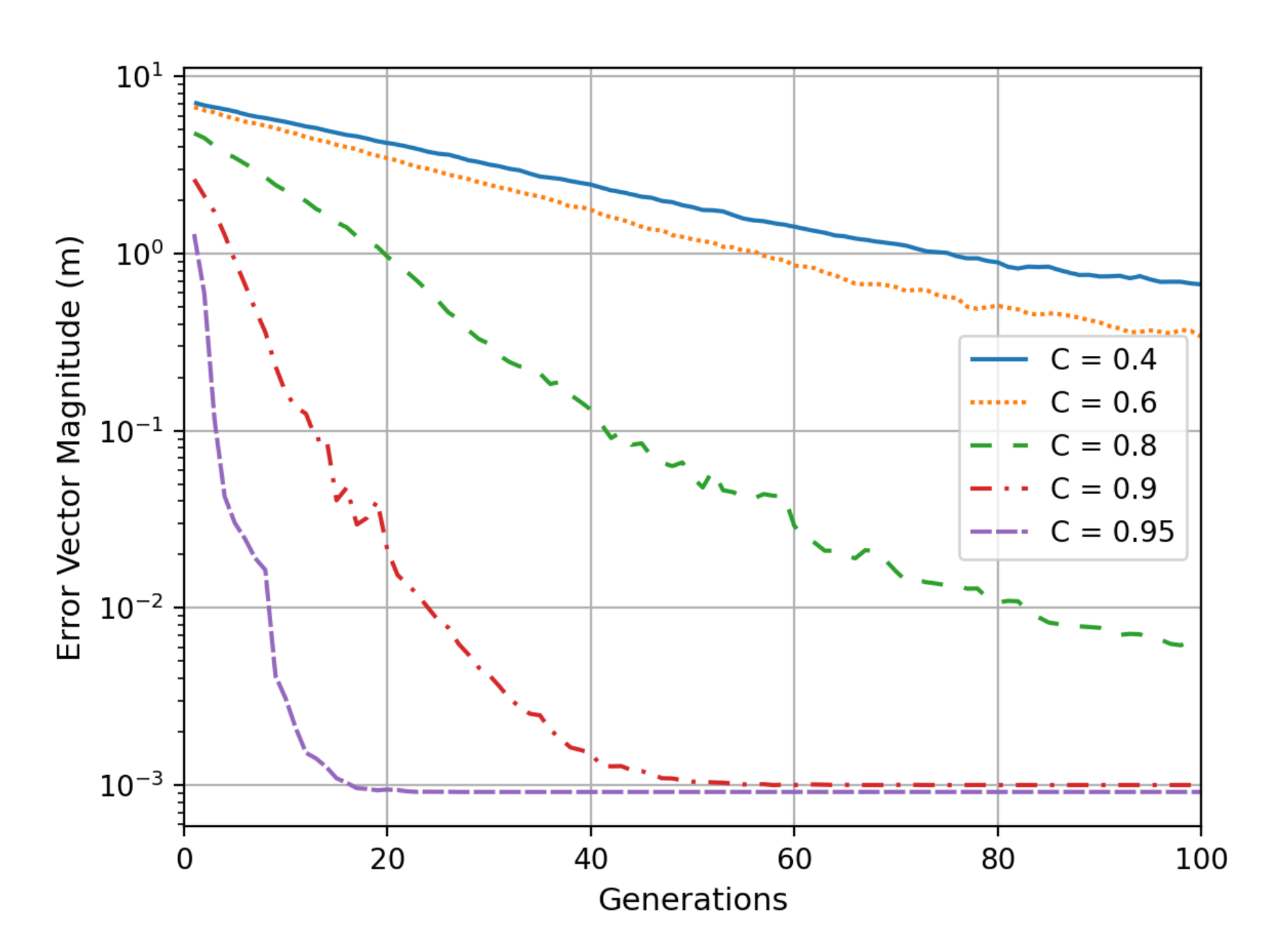}
	\caption{Simulated average algorithm localization accuracy (\ac{evm}) convergence plotted for multiple values of network connectivity. \ac{edm} estimates are generated with an array size of $N=10$ and a time-transfer bandwidth of \SI{40}{\mega\hertz}.}
	\label{fig:sim_evm_vs_c}
\end{figure}

\begin{figure}
	\centering
	\includegraphics[width=1\columnwidth]{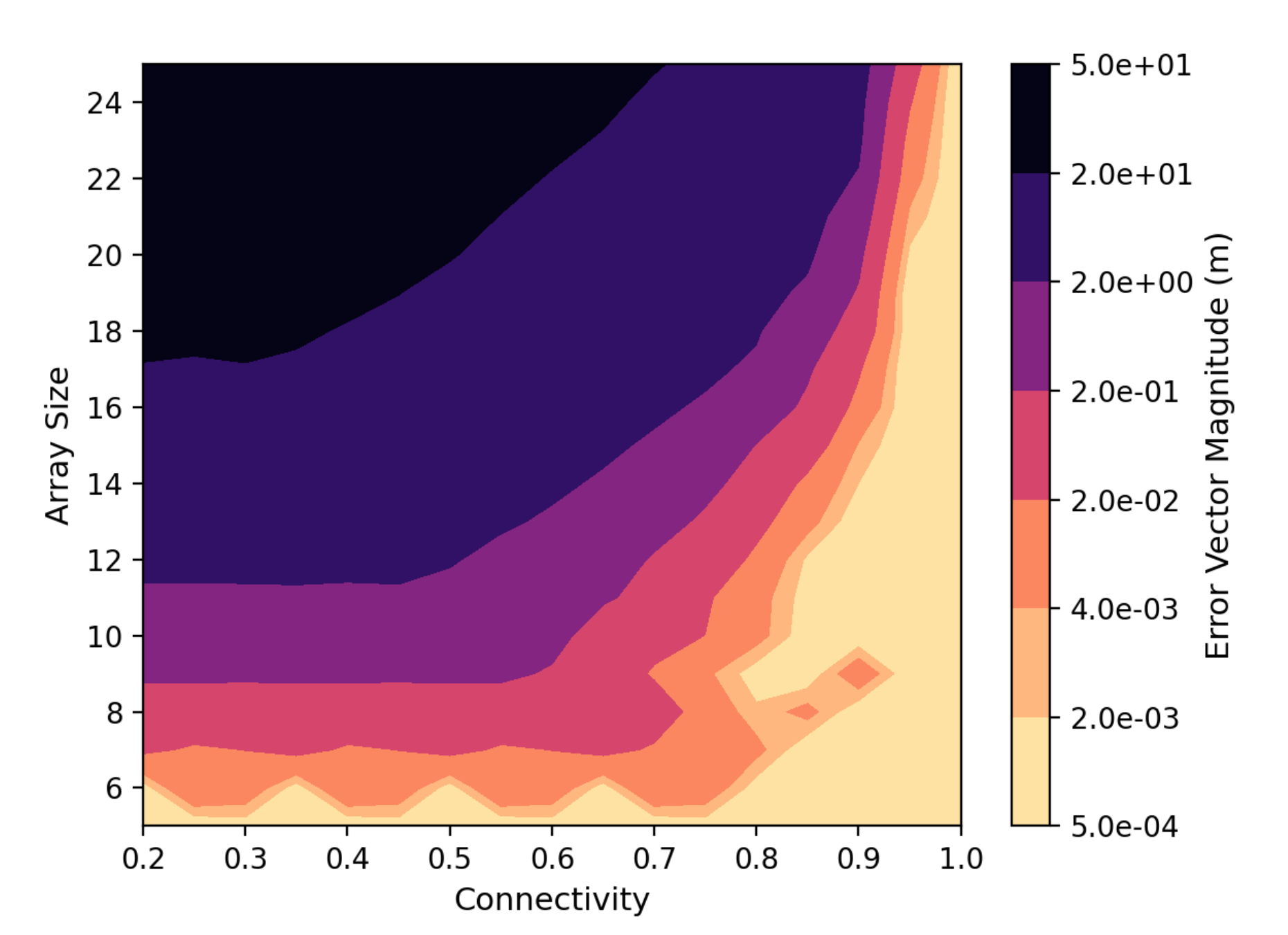}
	\caption{Contour plot describing the simulated average node localization error vector magnitude as a function of both array size and network connectivity. All \ac{edm} estimates were generated with a network \ac{snr} of \SI{34}{\dB} and a time-transfer tone separation of \SI{40}{\mega\hertz} to describe the error using the \ac{crlb}.}
	\label{fig:sim_evm_n_c}
\end{figure}

In order to verify the efficacy of the proposed algorithm, several Monte Carlo simulations were performed, varying array sizes and connectivity values. While \ac{snr} and time-transfer waveform bandwidth are also parameters of interest, their effect on ranging accuracy is clear due to their relationship to the \ac{crlb}, so for any system, the bandwidth or transmit power can be arbitrarily increased to improve localization results; thus, these parameters are not explored in simulation. For each Monte Carlo trial, element locations, distance estimation noise, and adjacency matrix structures were randomly generated. In Figures \ref{fig:sim_evm_vs_n}, \ref{fig:sim_evm_vs_c}, and \ref{fig:sim_evm_n_c}, data points represent the average of 250 independent runs of the proposed algorithm, with a harmonic mean \ac{snr} of \SI{34}{\dB} to emulate the noise response of the experimental environment to be described in Section \ref{sec:exp}. All simulations were performed using a \SI{2.445}{\giga\hertz} AMD EPYC 7763 128-core processor with \SI{493}{\giga\byte} of DDR4 RAM.

Figures \ref{fig:sim_evm_vs_n} and \ref{fig:sim_evm_vs_c} show the performance of the proposed localization algorithm over 100 generations for several values of array size and network connectivity, respectively. In Fig. \ref{fig:sim_evm_vs_n}, it is clear from the cases of $N=6,8,10$ that an increase in array size does not lead to an inability to converge to accurate node position estimates; in the larger cases of $N=15$ and $N=25$, millimeter-level accuracy is not achieved. It is evident because of the continuous decrease in the \ac{evm} that this is not due to premature convergence, but rather the limit enforced on the algorithm generations. Similarly, Fig. \ref{fig:sim_evm_vs_c} shows that as the connectivity of an array decreases, it is highly probable that it will take more generations to converge. Considering both the connectivity and array size, it can then be deduced that the rate of convergence of the algorithm is strongly dependent on the number of missing edges to be resolved, which is the number of variables in the optimization problem. In Fig. \ref{fig:meas_evm_vs_c}, additional simulation results are shown alongside the measurement results for the case of $N=6$, varying connectivity from minimally- to fully-connected.

In Figure \ref{fig:sim_evm_n_c}, the average simulated localization \ac{evm} is plotted as a function of both array size $N$ and connectivity ratio $c$. Each Monte Carlo instantiation was allowed to progress for 100 generations before its final \ac{evm} was evaluated; therefore, a high error ascribed to an area of the plot does not necessarily imply premature convergence, but rather the failure to converge in this number of function evaluations. Using the $\lambda/15$ metric described in Section \ref{sec:time}, the lightest region corresponding to smaller array sizes and especially high connectivities represents arrays that are able to converge within this generational limit to a precision that would enable beamforming above \SI{10}{\giga\hertz}. The following regions represent lower bounds of \SI{5}{\giga\hertz}, \SI{1}{\giga\hertz}, \SI{100}{\mega\hertz} and below this threshold, the positioning ability does not perform well enough to perform operations at frequencies of significant interest as the microwave and millimeter-wave bands.

\section{Localization Experiment}
\label{sec:exp}

\subsection{Experimental Configuration}
\label{sub_sec:setup}

\begin{figure*}
	\centering
	\includegraphics[width=0.9\textwidth]{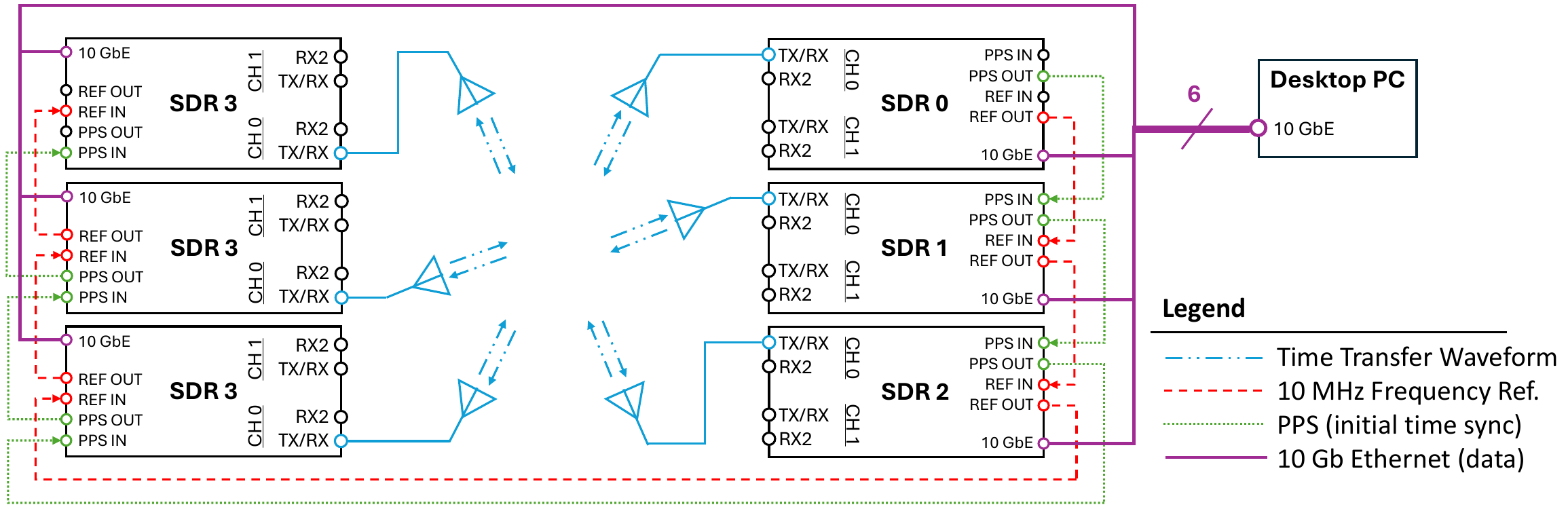}
	\caption{Time transfer system schematic for large-array internode range estimation. The output of the oscillator for \ac{sdr} 0 is designated as the frequency reference for all other nodes; node 0 also sends out a \ac{pps} signal upon startup to coarsely align the clocks, ensuring that the capture windows overlap for each pair of \acp{sdr}. A single desktop PC is connected to all \acp{sdr} via \SI{10}{\giga bit} Ethernet to control the time transfer and data collection processes using GNU Radio.}
	\label{fig:schematic}
\end{figure*}

\begin{table}[tb]
	\ifcsname hlon\endcsname%
		\color{blue}%
  	\fi%
	\caption{Experiment Parameters}
	\label{tab:exp_params}
	\begin{center}
  	\begin{tabularx}{\columnwidth}{p{0.43\linewidth}YY}
	
	\toprule[1pt]
	\textbf{Time Transfer Waveform} \\
	\midrule
	Parameter & Symbol & Value \\
	\midrule
	Waveform Type &  & \mbox{Pulsed\;Two-Tone} \\
	Carrier Frequency & $f_\mathrm{c}$ & \SI{2.1}{\giga\hertz} \\
	Tone Separation (Bandwidth) & $\beta_\mathrm{t}$ & \SI{40}{\mega\hertz} \\
	Rise/Fall Time & & \SI{50}{\nano\second} \\
	Pulse Duration & $\tau_\mathrm{p}$ & \SI{10.0}{\micro\second} \\
	Synchronization Epoch Duration &  & $\approx$\SI{700}{\milli\second} \\
	Resynchronization Interval &  & Software-Triggered \\
	Rx Sample Rate & $f^\mathrm{Rx}_\mathrm{s}$ & \SI{200}{\mega Sa/s} \\
	Tx Sample Rate & $f^\mathrm{Tx}_\mathrm{s}$ & \SI{400}{\mega Sa/s}* \\
	\midrule[1pt]
	\midrule[1pt]
	\textbf{Antenna Parameters}\\
	\midrule
	Parameter & Symbol & Value \\
	\midrule
	Gain &  & $1.75$\,dBi\\
	Bandwidth &  & \SI{1.7}{}--\SI{2.1}{\giga\hertz}\\
	Separation & & $\geq$\SI{45}{\centi\meter} \\
	\bottomrule[1pt]	
	\end{tabularx}
	\end{center}
	\footnotesize{* Digitally upsampled from \SI{200}{\mega Sa/s} to \SI{400}{\mega Sa/s} on device}
\end{table}

\begin{figure}
	\centering
	\includegraphics[width=1\columnwidth]{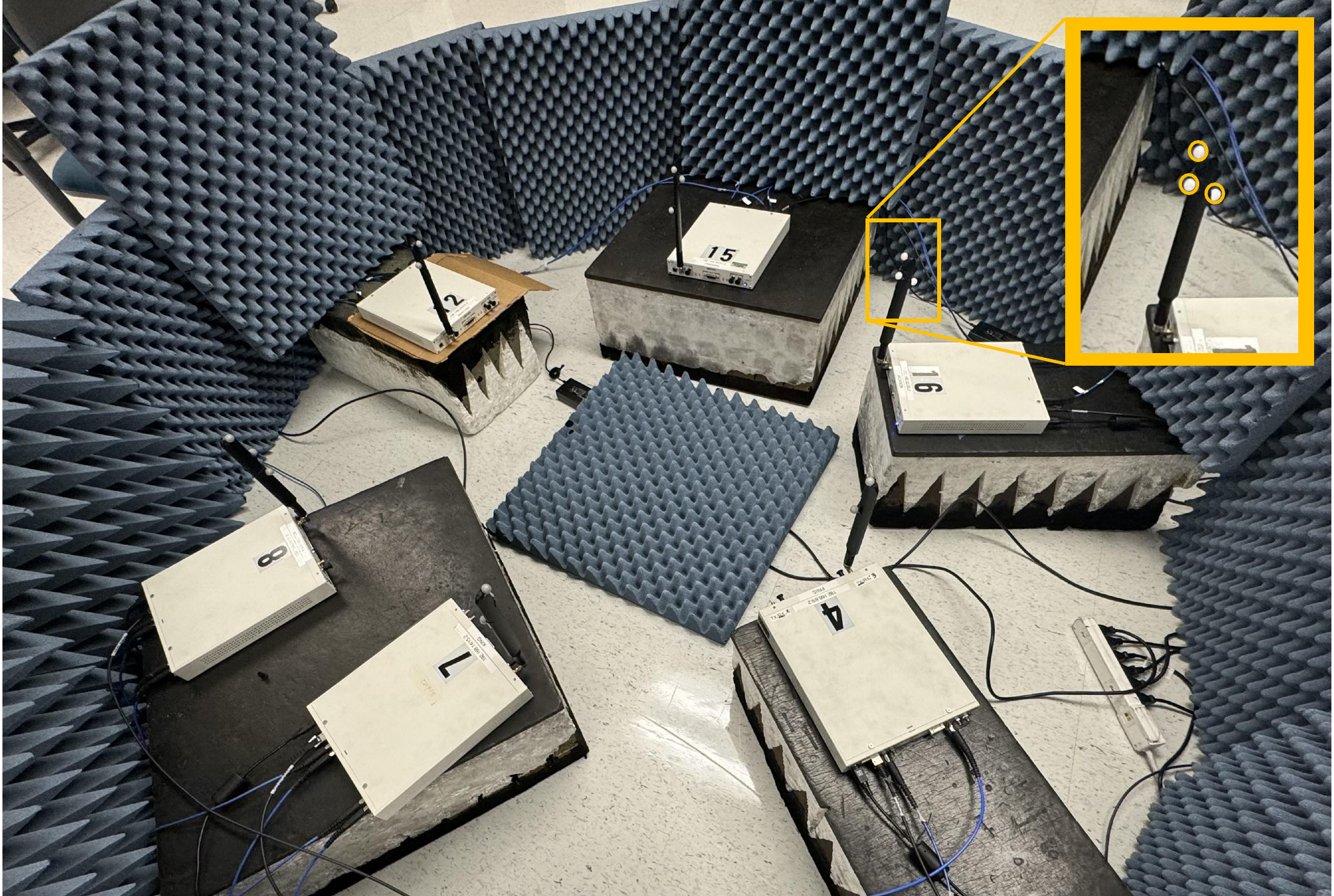}
	\caption{Image of the radio and antenna configuration for the experiment. Radios are directly connected to the dipole antennas, which are each fitted with optical reflectors for the motion capture system (see upper left inset).}
	\label{fig:setup}
\end{figure}

The ranging and localization experiment was conducted with a configuration of six nodes, performing ranging at multiple bandwidths; the system schematic is shown in Fig. \ref{fig:schematic}, and the setup is pictured in Fig. \ref{fig:setup}. The \ac{sdr}s used were Ettus Research \ac{usrp} X310's. Each radio was equipped with a UBX-160 daughterboard, running with a \SI{200}{\mega\hertz} clock and sampling rate of \SI{200}{\mega Sa/\second}. Because of the finite transmit and receive windowing allotted for the \acp{sdr}, a \ac{pps} link was connected to the radios; the \ac{pps} link was only used once upon startup of the measurement in order to coarsely align the clocks to within several ticks, so that times of reception would be ensured to arrive within the finite reception windows. The systems were also syntonized (aligned in frequency) by direct cabling, treating the oscillator of \ac{sdr}~0 as the frequency reference for all other nodes; note that wireless syntonization can be implemented in a future iteration~\cite{serge2020mwcl}. Each \ac{sdr} was also outfitted with a Pulse Electronics SPDA 24700/2700 dipole antenna with an operational bandwidth of \SI{1.7}{} to \SI{2.1}{\giga\hertz}.

In order to control the \acp{sdr}, each radio was linked over \SI{10}{\giga bit} Ethernet to a desktop computer; this computer contained a \SI{3}{\giga\hertz} Intel i7-9700 processor with \SI{16}{\giga\byte} RAM, running Ubuntu 20.04 Linux distribution. Installed on this computer, GNU Radio and the Ettus \ac{uhd} interacted with the \acp{sdr} to perform the time-transfer measurements. The \acp{sdr} were programmed in bursty mode in order to ensure high-accuracy timing estimation and to make use of the entire sampling bandwidth. Using the two-way time-transfer method described in Section \ref{sec:time}, two-tone waveform pulses were transmitted with a duration of \SI{10}{\micro\second} at a variety of bandwidths. The \acp{sdr} used a time-division multiple access approach to ensure that at any given time, only one node was transmitting and one was receiving, thus the \ac{pri} (synchronization epoch) was non-uniform, but on the order of hundreds of milliseconds.

To evaluate the accuracy of the algorithm applied to range estimates, the lab space was outfitted with six OptiTrack Flex13 cameras connected to an additional desktop computer running the OptiTrack Motive software, a motion capture system designed for sub-millimeter location tracking precision. As can be seen in Fig. \ref{fig:setup}, each antenna was fitted with three optical reflectors to triangulate the position of the center of the antennas. The antennas were then placed in an approximately circular pattern to ensure \ac{los} between each element, at equal elevations above ground level so that the system could be approximately constrained to reside in a plane, with dimension $m=2$. The OptiTrack system was also used to calibrate the system hardware and multipath environmental delays; with prior knowledge of the exact antenna locations, the \ac{rf} group delays through the system hardware were computed by subtracting the known times-of-flight from the total measured times.

For the measurements, the signal \ac{snr} between each link was estimated using the blind \ac{evd} technique \cite{sean2020taes,hamid2014}. The sample matrix $\textbf{S}$ is constructed as
\begin{equation}
	\label{eq:samp_matrix}
	\textbf{S} = 
	\begin{bmatrix}
	S_{1,1} & S_{1,2} & \cdots & S_{1,L} \\
	S_{2,1} & S_{2,2} & \cdots & S_{2,L} \\
	\vdots & \vdots & \ddots & \vdots \\
	S_{N,1} & S_{N,2} & \cdots & S_{N,L}
	\end{bmatrix}
\end{equation}
where the $L$ columns represent capture windows of $N$ samples each. The covariance matrix of $\boldsymbol{S}$ can then be constructed as
\begin{equation}
	\label{eq:samp_covariance}
	\textbf{R}_S = \frac{1}{N}\textbf{S}\textbf{S}^H
\end{equation}
where $(\cdot)^H$ represents the matrix Hermitian transpose operator. The matrix is then decomposed to find its eigenvalues. After sorting the eigenvalues in descending order of magnitude, $\gamma_1,\gamma_2,...,\gamma_L$, the noise power can be calculated as
\begin{equation}
	\label{eq:noise_power}
	P_\mathrm{n} = \frac{1}{L-1}\sum_{l=2}^L\gamma_l,
\end{equation}
and the signal power is given by
\begin{equation}
	\label{eq:sig_power}
	P_\mathrm{s} = \frac{\gamma_1-P_\mathrm{n}}{L}.
\end{equation}
The \ac{snr} was calculated for each link in this manner, and tuned to be within a range of \SI{32}{}--\SI{36}{\dB}, with a harmonic mean value of \SI{34}{\dB}.

\subsection{Experimental Results}
\label{sub_sec:results}

\begin{figure}
	\centering
	\includegraphics[width=1\columnwidth]{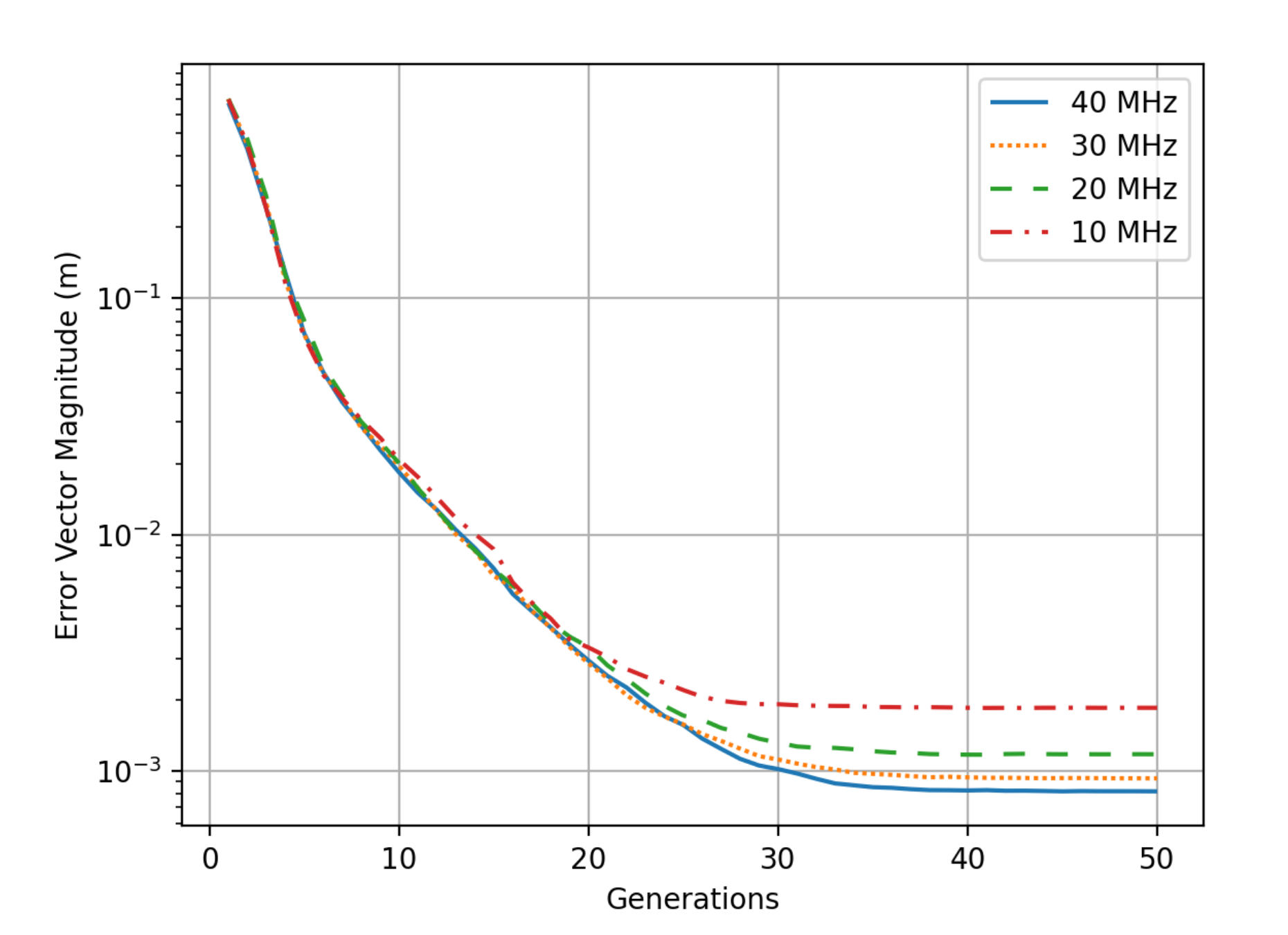}
	\caption{Average algorithm localization accuracy (\ac{evm}) convergence vs. tone separation of the two-tone range estimation waveform. For each \ac{edm} estimate collected, the algorithm is initialized with a randomly-generated, minimally-connected ($c=0.8$ for $N=6$) adjacency matrix; \acp{edm} were collected at a harmonic mean link \ac{snr} of \SI{34}{\dB} and a carrier frequency of $f_\mathrm{c}=$\SI{2.1}{\giga\hertz}.}
	\label{fig:meas_evm_vs_b}
\end{figure}

\begin{figure}
	\centering
	\includegraphics[width=1\columnwidth]{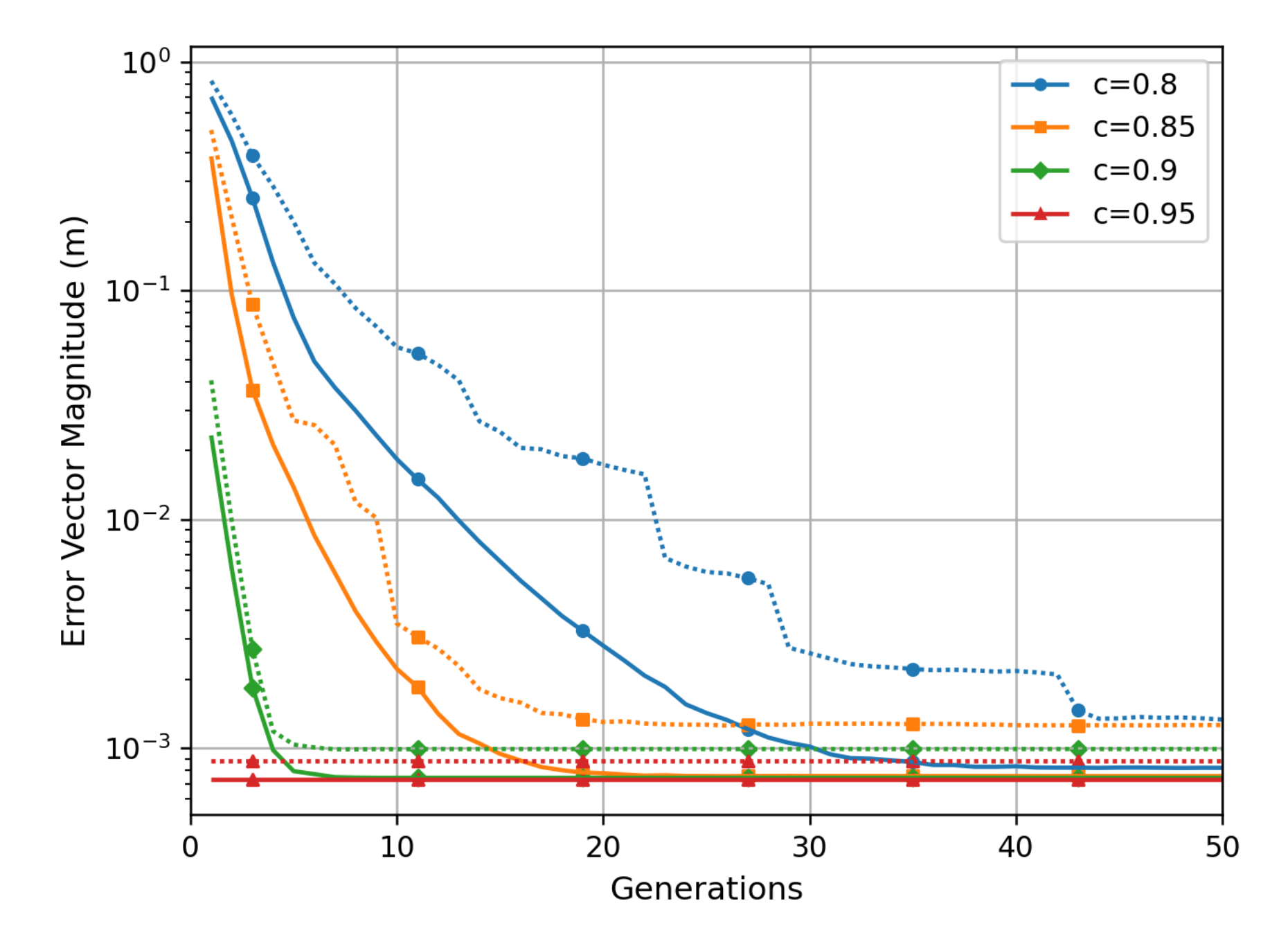}
	\caption{Average algorithm localization accuracy (\ac{evm}) convergence vs. connectivity of the network adjacency matrix for measured (solid) and simulated (dotted) results. \acp{edm} estimates were collected with a two-tone waveform tone separation of \SI{40}{\mega\hertz} at a harmonic mean link \ac{snr} of \SI{34}{\dB} and a carrier frequency of $f_\mathrm{c}=$\SI{2.1}{\giga\hertz}.}
	\label{fig:meas_evm_vs_c}
\end{figure}

While collecting the range estimates, each synchronization epoch occurred over several hundred milliseconds in an environment with an average link \ac{snr} of \SI{34}{\dB} and performed estimation for approximately ten minutes. With these parameters fixed, the bandwidth was varied from \SI{10}{} to \SI{40}{\mega\hertz} and for each test case, 250 estimates of the system \ac{edm} were collected in order to evaluate the average ranging precision and performance of the proposed algorithm. Additionally, links between nodes were artificially severed in post-processing by randomizing the adjacency matrix in order to further evaluate the algorithm performance versus each connectivity ratio. Results for the sweep of bandwidth and connectivity are shown in Figs. \ref{fig:meas_evm_vs_b} and \ref{fig:meas_evm_vs_c}, respectively. In Fig. \ref{fig:meas_evm_vs_b}, each curve represents a measurement taken with a distinct tone separation, ranging from \SI{10}{}--\SI{40}{\mega\hertz}. Each curve represents the average \ac{evm} of approximately 250 \ac{edm} estimates taken across 50 generations of the genetic algorithm, initialized with a population size of 200. It is clear that the bandwidth variation does not noticably affect the accuracy of convergence, especially in the early generations; however, as expected and stated in Section \ref{sub_sec:mc_sims}, due to the inverse relationship between $\sigma_\mathrm{d}^2$ and bandwidth, increasing the bandwidth of the waveform results in improved localization precision.

In Fig. \ref{fig:meas_evm_vs_c}, the same number of estimates are evaluated across 50 generations with the same population size, shown by the solid lines. Simulated results, taken from the same random distribution described in Section \ref{sec:sim}, are shown alongside the measurements as dotted lines, with colors matching the measurements of the same connectivity. This time, they are all taken from the \SI{40}{\mega\hertz} measurement, and instead the connectivity of the genetic algorithm adjacency matrix is varied for each estimate; the average performance at each generation is plotted for each value of connectivity. Here, we see that the final value to which the algorithm converges is affected by less than a tenth of a millimeter. However, since there are more variables to optimize in the more sparsely-connected cases, the algorithm takes more generations to achieve convergence as expected. It is also notable that the measured results seem to outperform the simulation. This is likely due to the optimal layout of the experimental setup, with the nodes arranged in a roughly circular pattern so that all internode links have a similar \ac{snr}, so that eliminating certain links does not have a dramatic affect on the overall network \ac{snr}; in the simulations, randomized layouts with randomized connectivity could cut connections with \acp{snr} much higher than the average value, resulting in higher error at convergence. In the most sparsely-connected case, we are still able to achieve sub-millimeter accuracy, with a mean \ac{evm} of \SI{0.82}{\milli\meter}.

\subsection{Discussion}
\label{sub_sec:disc}

As discussed in the introduction, one of the main benefits of the distributed array technique is the decentralized aspect; because no single point is designated as the origin, we are able to evaluate positioning error on a node-by-node basis. Additionally, the distances are equally weighted in the optimization problem, so errors will not propagate through the array as they would in a centralized system as we attempt to localize elements furthest from the primary node. Another proven benefit of the demonstrated method is that we are clearly able to resolve missing connections with a negligible change in system positioning error using a spectrally-sparse ranging approach. This has positive implications for practical applications, where system constraints may dictate limited connectivity, and the spectral efficiency of the two-tone waveform provides the opportunity to perform simultaneous internode or targeted operations by minimizing the frequency allocation needed for positioning and synchronization.

While genetic algorithms are able to explore an arbitrarily wide search space, making them good candidates for solving non-convex problems, there are two underlying issues with the approach, which can be observed in Fig. \ref{fig:meas_evm_vs_c}. First, the number of function evaluations required to converge can be large; second, convergence to the global minimum is not entirely guaranteed. These issues are exacerbated as the number of variables involved in the cost function increases due to matrix sparsity, as established in Section \ref{sec:sim}. In order to decrease the probability of premature convergence to any of the set of undesirable local minima, the algorithm population size should be increased; however, such an increase also increases the computational time, which could quickly become prohibitive for some applications. 

If the array element positions are fixed, or if they are designed to be reconfigurable but not in relative motion during operation, then computational time may not be problematic. However, in many practical systems such as satellite or vehicle-to-vehicle communications, it is likely desirable to compute and track element positions quickly when the assumption of fixed relative location cannot be made. In this paper, we introduce the concept of decentralized genetic localization, though the actual computation relies on a central computing node. It may prove beneficial to distribute the needed computation among the array nodes and localize the system in smaller subsets of elements to decrease computation time. Other algorithms may exist that solve the \ac{edmcp} more efficiently, but we have shown that especially for a small degree of incompleteness, a \ac{ga} approach can serve to effectively localize a set of distributed elements.

\section{Conclusion}
\label{sec:conclusion}

In this paper, we demonstrated a distributed antenna array system capable of element localization in a sparsely-connected network topology. We achieved a precision of \SI{0.82}{\milli\meter} with an average link \ac{snr} of \SI{34}{\dB} through the implementation of an evolutionary algorithm with an \ac{mds}-based cost function. This work represents a significant step toward fully-decentralized function distributed networks and distributed phased array operations, achieving localization accuracy that would support beamforming capabilities up to \SI{24.3}{\giga\hertz}.

\bibliography{mdula_refs.bib}

\begin{thebibliography}{10}
\providecommand{\url}[1]{#1}
\csname url@samestyle\endcsname
\providecommand{\newblock}{\relax}
\providecommand{\bibinfo}[2]{#2}
\providecommand{\BIBentrySTDinterwordspacing}{\spaceskip=0pt\relax}
\providecommand{\BIBentryALTinterwordstretchfactor}{4}
\providecommand{\BIBentryALTinterwordspacing}{\spaceskip=\fontdimen2\font plus
\BIBentryALTinterwordstretchfactor\fontdimen3\font minus \fontdimen4\font\relax}
\providecommand{\BIBforeignlanguage}[2]{{%
\expandafter\ifx\csname l@#1\endcsname\relax
\typeout{** WARNING: IEEEtran.bst: No hyphenation pattern has been}%
\typeout{** loaded for the language `#1'. Using the pattern for}%
\typeout{** the default language instead.}%
\else
\language=\csname l@#1\endcsname
\fi
#2}}
\providecommand{\BIBdecl}{\relax}
\BIBdecl

\bibitem{yuan2011}
L.~Yuan, G.~Zheng, and X.~Li, ``Research on theory and technology of distributed mimo radar systems,'' in \emph{Proceedings of 2011 IEEE CIE International Conference on Radar}, vol.~1, 2011, pp. 87--90.

\bibitem{hasnain2018}
S.~N. Hasnain, R.~Stephan, M.~Brachvogel, M.~Meurer, and M.~A. Hein, ``Performance of distributed antenna sub-arrays for robust satellite navigation in automotive applications,'' \emph{European Journal of Navigation}, 2018.

\bibitem{tagliaferri2021}
D.~Tagliaferri, M.~Rizzi, S.~Tebaldini, M.~Nicoli, I.~Russo, C.~Mazzucco, A.~V. Monti-Guarnieri, C.~M. Prati, and U.~Spagnolini, ``Cooperative synthetic aperture radar in an urban connected car scenario,'' in \emph{2021 1st IEEE International Online Symposium on Joint Communications \& Sensing (JC\&S)}, 2021, pp. 1--4.

\bibitem{nasa2020tax}
``2020 {NASA} technology taxonomy,'' National Aeronautics and Space Administration, Tech. Rep. HQ-E-DAA-TN76545, Jan. 2020.

\bibitem{nanzer2017tmtt}
J.~A. Nanzer, R.~L. Schmid, T.~M. Comberiate, and J.~E. Hodkin, ``Open-loop coherent distributed arrays,'' \emph{IEEE Transactions on Microwave Theory and Techniques}, vol.~65, no.~5, pp. 1662--1672, 2017.

\bibitem{nanzer2021tmtt}
J.~A. Nanzer, S.~R. Mghabghab, S.~M. Ellison, and A.~Schlegel, ``Distributed phased arrays: Challenges and recent advances,'' \emph{IEEE Transactions on Microwave Theory and Techniques}, vol.~69, no.~11, pp. 4893--4907, 2021.

\bibitem{zafari2015journal}
F.~Zafari, I.~Papapanagiotou, and K.~Christidis, ``Microlocation for internet-of-things-equipped smart buildings,'' \emph{IEEE Internet of Things Journal}, vol.~3, no.~1, pp. 96--112, 2015.

\bibitem{jovanovic2023}
M.~D. Jovanovic and S.~M. Djosic, ``Analysis of indoor localization techniques,'' in \emph{2023 58th International Scientific Conference on Information, Communication and Energy Systems and Technologies (ICEST)}, 2023, pp. 219--222.

\bibitem{jovanovic2018}
M.~Jovanovic, S.~Djosic, and I.~Stojanovi{\'c}, ``Analysis of ultra wideband localization techniques,'' in \emph{XIV Int. SAUM Conf.}, 2018.

\bibitem{paul2009}
A.~S. Paul and E.~A. Wan, ``Rssi-based indoor localization and tracking using sigma-point kalman smoothers,'' \emph{IEEE Journal of selected topics in signal processing}, vol.~3, no.~5, pp. 860--873, 2009.

\bibitem{barsocchi2009}
P.~Barsocchi, S.~Lenzi, S.~Chessa, and G.~Giunta, ``A novel approach to indoor rssi localization by automatic calibration of the wireless propagation model,'' in \emph{VTC Spring 2009-IEEE 69th Vehicular Technology Conference}.\hskip 1em plus 0.5em minus 0.4em\relax IEEE, 2009, pp. 1--5.

\bibitem{bisio2014}
I.~Bisio, M.~Cerruti, F.~Lavagetto, M.~Marchese, M.~Pastorino, A.~Randazzo, and A.~Sciarrone, ``A trainingless wifi fingerprint positioning approach over mobile devices,'' \emph{IEEE Antennas and Wireless Propagation Letters}, vol.~13, pp. 832--835, 2014.

\bibitem{loyez2015}
C.~Loyez, M.~Bocquet, C.~Lethien, and N.~Rolland, ``A distributed antenna system for indoor accurate wifi localization,'' \emph{IEEE Antennas and Wireless Propagation Letters}, vol.~14, pp. 1184--1187, 2015.

\bibitem{kravets2024}
I.~Kravets, O.~Kapshii, O.~Shuparskyy, and A.~Luchechko, ``A new parametric three stage weighted least squares algorithm for tdoa-based localization,'' \emph{IEEE Access}, 2024.

\bibitem{alsamhi2019}
\BIBentryALTinterwordspacing
S.~H. Alsamhi, O.~Ma, M.~S. Ansari, and Q.~Meng, ``Greening internet of things for greener and smarter cities: a survey and future prospects,'' \emph{Telecommunication Systems}, vol.~72, no.~4, pp. 609--632, Dec 2019. [Online]. Available: \url{https://doi.org/10.1007/s11235-019-00597-1}
\BIBentrySTDinterwordspacing

\bibitem{bhat2022}
\BIBentryALTinterwordspacing
S.~J. Bhat and S.~K~V, ``A localization and deployment model for wireless sensor networks using arithmetic optimization algorithm,'' \emph{Peer-to-Peer Networking and Applications}, vol.~15, no.~3, pp. 1473--1485, May 2022. [Online]. Available: \url{https://doi.org/10.1007/s12083-022-01302-x}
\BIBentrySTDinterwordspacing

\bibitem{deutschmann2024}
B.~J. Deutschmann, C.~Nelson, M.~Henriksson, G.~Marti, A.~Kosasih, N.~Tervo, E.~Leitinger, and F.~Tufvesson, ``Accurate direct positioning in distributed mimo using delay-doppler channel measurements,'' \emph{arXiv preprint arXiv:2404.15936}, 2024.

\bibitem{li2020basestation}
Y.~Li, Z.~Zhang, L.~Wu, J.~Dang, and P.~Liu, ``5g communication signal based localization with a single base station,'' in \emph{2020 IEEE 92nd Vehicular Technology Conference (VTC2020-Fall)}, 2020, pp. 1--5.

\bibitem{zhang2022}
D.~Zhang, J.~Han, G.~Cheng, and M.-H. Yang, ``Weakly supervised object localization and detection: A survey,'' \emph{IEEE Transactions on Pattern Analysis and Machine Intelligence}, vol.~44, no.~9, pp. 5866--5885, 2022.

\bibitem{zafari2019survey}
F.~Zafari, A.~Gkelias, and K.~K. Leung, ``A survey of indoor localization systems and technologies,'' \emph{IEEE Communications Surveys \& Tutorials}, vol.~21, no.~3, pp. 2568--2599, 2019.

\bibitem{jason2023tmtt}
J.~M. Merlo, S.~R. Mghabghab, and J.~A. Nanzer, ``Wireless picosecond time synchronization for distributed antenna arrays,'' \emph{IEEE Transactions on Microwave Theory and Techniques}, vol.~71, no.~4, pp. 1720--1731, 2023.

\bibitem{anton2020mwcl}
A.~Schlegel, S.~M. Ellison, and J.~A. Nanzer, ``A microwave sensor with submillimeter range accuracy using spectrally sparse signals,'' \emph{IEEE Microwave and Wireless Components Letters}, vol.~30, no.~1, pp. 120--123, 2020.

\bibitem{ahona2023eurad}
A.~Bhattacharyya, J.~M. Merlo, and J.~A. Nanzer, ``A dual-carrier linear-frequency modulated waveform for high-accuracy localization in distributed antenna arrays,'' in \emph{2023 20th European Radar Conference (EuRAD)}, 2023, pp. 331--334.

\bibitem{dula2024wamicon}
M.~J. Dula, N.~Shandi, and J.~A. Nanzer, ``Decentralized localization of distributed phased array elements using high-accuracy ranging and multidimensional scaling,'' in \emph{2024 IEEE Wireless and Microwave Technology Conference (WAMICON)}, 2024, pp. 1--4.

\bibitem{naim2024tcom}
N.~Shandi, J.~M. Merlo, and J.~A. Nanzer, ``Decentralized picosecond synchronization for distributed wireless systems,'' \emph{IEEE Transactions on Communications}, 2024.

\bibitem{richards2014rsp}
M.~A. Richards, \emph{Fundamentals of Radar Signal Processing}.\hskip 1em plus 0.5em minus 0.4em\relax McGraw-Hill Education, 2014, vol.~1.

\bibitem{nanzer2016tap}
J.~A. Nanzer and M.~D. Sharp, ``On the estimation of angle rate in radar,'' \emph{IEEE Transactions on Antennas and Propagation}, vol.~65, no.~3, pp. 1339--1348, 2016.

\bibitem{moddemeijer1991}
R.~Moddemeijer, ``On the determination of the position of extrema of sampled correlators,'' \emph{IEEE Transactions on Signal Processing}, vol.~39, no.~1, pp. 216--219, 1991.

\bibitem{serge2022mwcl}
S.~R. Mghabghab and J.~A. Nanzer, ``Microwave ranging via least-squares estimation of spectrally sparse signals in software-defined radio,'' \emph{IEEE Microwave and Wireless Components Letters}, vol.~32, no.~2, pp. 161--164, 2022.

\bibitem{torgerson1952}
W.~S. Torgerson, ``Multidimensional scaling: I. theory and method,'' \emph{Psychometrika}, vol.~17, no.~4, pp. 401--419, 1952.

\bibitem{gower1982}
J.~C. Gower, ``Euclidean distance geometry,'' \emph{Mathematical Scientist}, vol.~7, pp. 1--14, 1982.

\bibitem{dokmanic2015}
I.~Dokmanic, R.~Parhizkar, J.~Ranieri, and M.~Vetterli, ``Euclidean distance matrices: Essential theory, algorithms, and applications,'' \emph{IEEE Signal Processing Magazine}, vol.~32, no.~6, pp. 12--30, 2015.

\bibitem{keshavan2009noisy}
R.~Keshavan, A.~Montanari, and S.~Oh, ``Matrix completion from noisy entries,'' \emph{Advances in neural information processing systems}, vol.~22, 2009.

\bibitem{rus2004}
D.~Moore, J.~Leonard, D.~Rus, and S.~Teller, ``Robust distributed network localization with noisy range measurements,'' in \emph{Proceedings of the 2nd international conference on Embedded networked sensor systems}, 2004, pp. 50--61.

\bibitem{fang2012}
\BIBentryALTinterwordspacing
H.~ren Fang and D.~P. O'Leary, ``Euclidean distance matrix completion problems,'' \emph{Optimization Methods and Software}, vol.~27, no. 4-5, pp. 695--717, 2012. [Online]. Available: \url{https://doi.org/10.1080/10556788.2011.643888}
\BIBentrySTDinterwordspacing

\bibitem{alfakih1999}
A.~Y. Alfakih, A.~Khandani, and H.~Wolkowicz, ``Solving euclidean distance matrix completion problems via semidefinite programming,'' \emph{Computational optimization and applications}, vol.~12, pp. 13--30, 1999.

\bibitem{cetin1993}
B.~Cetin, J.~Burdick, and J.~Barhen, ``Global descent replaces gradient descent to avoid local minima problem in learning with artificial neural networks,'' in \emph{IEEE International Conference on Neural Networks}, 1993, pp. 836--842 vol.2.

\bibitem{pymoo}
J.~Blank and K.~Deb, ``Pymoo: Multi-objective optimization in python,'' \emph{IEEE Access}, vol.~8, pp. 89\,497--89\,509, 2020.

\bibitem{haupt2004}
R.~L. Haupt and S.~E. Haupt, \emph{Practical genetic algorithms}.\hskip 1em plus 0.5em minus 0.4em\relax John Wiley \& Sons, 2004.

\bibitem{serge2020mwcl}
S.~R. Mghabghab, S.~M. Ellison, and J.~A. Nanzer, ``Open-loop distributed beamforming using wireless phase and frequency synchronization,'' \emph{IEEE Microwave and Wireless Components Letters}, vol.~32, no.~3, pp. 234--237, 2022.

\bibitem{sean2020taes}
S.~M. Ellison and J.~A. Nanzer, ``High-accuracy multinode ranging for coherent distributed antenna arrays,'' \emph{IEEE Transactions on Aerospace and Electronic Systems}, vol.~56, no.~5, pp. 4056--4066, 2020.

\bibitem{hamid2014}
M.~Hamid, N.~Bj{\"o}rsell, and S.~B. Slimane, ``Sample covariance matrix eigenvalues based blind snr estimation,'' in \emph{2014 IEEE International Instrumentation and Measurement Technology Conference (I2MTC) Proceedings}.\hskip 1em plus 0.5em minus 0.4em\relax IEEE, 2014, pp. 718--722.

\end{thebibliography}
\bibliographystyle{IEEEtran}
\end{document}